\documentclass[12pt,epsfig]{article}
\pdfoutput=1
\usepackage{amsmath,amssymb}
\usepackage{graphicx}
\usepackage[FIGTOPCAP]{subfigure}
\def\beq{\begin{equation}}
\def\eeq#1{\label{#1}\end{equation}}
\def\eeqn{\end{equation}}
\def\beqa{\begin{eqnarray}}
\def\eeqa#1{\label{#1}\end{eqnarray}}
\def\eeqan{\end{eqnarray}}

\usepackage{tikz}
\usetikzlibrary{arrows,shapes}
\usetikzlibrary{trees}
\usetikzlibrary{matrix,arrows} 				
\usetikzlibrary{positioning}				
\usetikzlibrary{calc,through}				
\usetikzlibrary{decorations.pathreplacing}  
\usepackage{pgffor}							

\usetikzlibrary{decorations.pathmorphing}	
\usetikzlibrary{decorations.markings}
\usetikzlibrary{snakes}

\tikzset{
    vector/.style={decorate, decoration={snake}, draw},
	provector/.style={decorate, decoration={snake,amplitude=2.5pt}, draw},
	antivector/.style={decorate, decoration={snake,amplitude=-2.5pt}, draw},
    fermion/.style={draw, postaction={decorate},
        decoration={markings,mark=at position .55 with {\arrow[draw]{>}}}},
    fermionbar/.style={draw, postaction={decorate},
        decoration={markings,mark=at position .55 with {\arrow[draw=black]{<}}}},
    fermionnoarrow/.style={draw},
    gluon/.style={decorate, draw,decoration={coil,amplitude=4pt, segment length=6pt}, line width=1},
    scalar/.style={dashed,draw, postaction={decorate},
        decoration={markings,mark=at position .55 with {\arrow[draw]{>}}}},
    scalarbar/.style={dashed,draw, postaction={decorate},
        decoration={markings,mark=at position .55 with {\arrow[draw]{<}}}},
    scalarnoarrow/.style={dash pattern = on 6 pt off 3 pt,draw},
    electron/.style={draw, postaction={decorate},
        decoration={markings,mark=at position .55 with {\arrow[draw]{>}}}},
	bigvector/.style={decorate, decoration={snake,amplitude=4pt}, draw},
	vectorscalar/.style={loosely dotted,draw, postaction={decorate}},
}

\def\lsim{\mathrel{\rlap{\lower4pt\hbox{\hskip1pt$\sim$}}
    \raise1pt\hbox{$<$}}}                
\def\gsim{\mathrel{\rlap{\lower4pt\hbox{\hskip1pt$\sim$}}
    \raise1pt\hbox{$>$}}}                

\def\stacksymbols #1#2#3#4{\def\theguybelow{#2}
    \def\vp{\lower#3pt}
    \def\sp{\baselineskip0pt\lineskip#4pt}
    \mathrel{\mathpalette\intermediary#1}}

\def\intermediary#1#2{\vp\vbox{\sp
     \everycr={}\tabskip0pt
     \halign{$\mathsurround0pt#1\hfil##\hfil$\crcr#2\crcr
              \theguybelow\crcr}}}

\def\gsim{\stacksymbols{>}{\sim}{2.5}{.2}}
\def\lsim{\stacksymbols{<}{\sim}{2.5}{.2}}

\begin{document}

\begin{titlepage}

\begin{center}

{\huge \bf A 130\,GeV Gamma Ray Signal}

\vskip.3cm

{\Huge \bf  from Supersymmetry}
\end{center}
\vskip1cm

\begin{center}



{\bf Bibhushan Shakya} \\
\end{center}
\vskip 8pt

\begin{center}
	{\it Laboratory for Elementary Particle Physics,\\
	     Cornell University, Ithaca, NY 14853, USA } \\

\vspace*{0.4cm}

{\tt  bs475@cornell.edu}
\end{center}

\vglue 0.3truecm

\vskip 0.5cm

\abstract
\noindent{The viability of neutralino dark matter as an explanation of the 130 GeV gamma ray signal from the Galactic Center recently observed by the Fermi Large Area Telescope is examined. It is found that the signal can be compatible with a sharp feature from internal bremsstrahlung from a mostly bino dark matter particle of mass around 145\,GeV, augmented by a contribution from annihilation into gamma+Z via a small wino admixture. This scenario circumvents the problematic overproduction of lower energy continuum photons  that plague line interpretations of this signal. Sleptons approximately degenerate in mass with the neutralino are required to enhance the internal bremsstrahlung feature.}

\vspace{1cm}


\end{titlepage}

\section{Motivation}

Several analyses (\cite{130weniger}\cite{130finkbeiner}\cite{130tempel}\cite{brem}) have recently confirmed the presence of a sharp feature, incompatible with conventional astrophysics, at an energy of approximately 130\,GeV in the gamma ray spectrum towards the Galactic Center in the data gathered by the Fermi Large Area Telescope (LAT). While the possibility that this might be an instrumental effect or a product of nonconventional astrophysics still exists, such a feature has long been earmarked as a ``smoking gun" signature of dark matter annihilation in the galaxy; this tantalizing interpretation has therefore generated significant excitement. 

Assuming a dark matter origin, the signal is best fit by a 130\,GeV dark matter particle pair-annihilating into photons with an annihilation cross section of $\langle\sigma v\rangle_{\gamma\gamma}=1.27\times10^{-27}$cm$^3$s$^{-1}$, assuming an Einasto profile for the dark matter distribution \cite{130weniger}. From a particle physics point of view, this scenario poses two major problems. First, since dark matter is not expected to couple directly to photons, annihilation to a photon pair must occur via a loop (see \cite{loop1},\cite{loop2} for a full calculation of this process in supersymmetry); for a thermal relic, this loop-suppressed cross section is generally too small to produce the signal observed by  Fermi. Second, even if this cross section can be made large enough, tree-level annihilation to particles that mediate the photon pair production process should produce a large continuum of photons at lower energies, which is not seen in the  Fermi data \cite{continuum}\cite{continuum2}\cite{continuum3}\cite{wocontinuum}. These considerations have been shown to rule out the most promising and the most studied dark matter candidate, the lightest neutralino in supersymmetry, as an explanation of this line signal \cite{continuum}\cite{continuum2}\cite{continuum3}.  

However, a monochromatic line signal is not the only possibility that can explain this feature; it is well-known that internal bremsstrahlung (hereafter IB) --  the production of a photon in conjunction with the leading annihilation channel into charged particles -- can also give sharp spectral features in the $\gamma$ ray spectrum close to the dark matter mass \cite{ibdiscussion}\cite{ibheavyneutralino}\cite{robust}. The first evidence of  the 130 GeV feature was, in fact, found in searches for IB signals in the Fermi LAT data \cite{brem}, where a fit significance of $3.2(4.6)\sigma$ excess with (without) the look elsewhere effect was found for an IB signal corresponding to annihilation of a dark matter particle of mass around 150\,GeV. While an IB signal is broader and gives a poorer fit to Fermi data than a line signal, it already appears at tree-level and can potentially overcome the constraints from signal strength and overproduction of continuum photons that plague line signals. 

The purpose of this paper is to examine whether IB from neutralino dark matter annihilation can provide a viable explanation of the 130\,GeV signal observed by Fermi. The possibility of explaining the feature with IB has been mentioned in several papers
, but an explicit treatment within supersymmetry is still lacking. While the toy model employed in \cite{brem} is very similar in several respects, there are several crucial differences, such as the availability of several IB channels and significant contributions from the $\gamma\gamma$ and $\gamma Z$ lines, and the supersymmetric parameter space is not treated in detail. Given the overwhelming popularity of the neutralino as a dark matter candidate, a more careful study is therefore desirable. 

This paper is organized as follows. Section 2 discusses continuum constraints and IB in the context of supersymmetry. Section 3 contains details and discussions of the scans, a few benchmark points, and their fit to the Fermi data. Section 4 discusses relic density and other relevant dark matter considerations. The main results are summarized in Section 5.     



\section{Neutralino Dark Matter and Internal Bremsstrahlung}

In the Minimal Supersymmetric Standard Model (MSSM), neutralino dark matter is an admixture of the bino, neutral wino, and the two neutral Higgsinos, and its composition determines the strength of the gamma ray signal and the relative size of the continuum. Table \ref{tab:limits} lists the approximate annihilation cross sections into $\gamma\gamma$, $\gamma Z$, and the total cross section for a dark matter particle of mass 130\,GeV in the three limits (taken from \cite{continuum}).

\begin{table}[h]
\begin{center}
\begin{tabular}{|l|c|c|c|c|} \hline
Model & $\sigma_{\gamma\gamma}v$ (cm$^3$s$^{-1}$) & $\sigma_{\gamma Z}v$ (cm$^3$s$^{-1}$) & $\sigma_{total}v$ (cm$^3$s$^{-1}$) & $R^{th}$\\
\hline
Bino & $\sim10^{-30}$ & $\sim10^{-31}$ & $\sim10^{-27}$  & $\sim1000$ \\
Wino & 2.5$\times10^{-27}$ & 1.4$\times10^{-26}$ & 4$\times10^{-24}$ & 210 \\
Higgsino & 1.1$\times10^{-28}$ & 3.7$\times10^{-28}$  & 4.2$\times10^{-25}$  & 710 \\
\hline
\end{tabular}
\caption{Annihilation cross sections for various neutralino constituents. The bino cross sections are with slepton masses at 200\,GeV. $R^{th}$ represents the size of the continuum relative to the line signal \cite{continuum}.}
\label{tab:limits}
\end{center}
\end{table}

Recall that the best fit to the Fermi data requires an annihilation cross section of $\sigma_{\gamma\gamma}v=1.27\times 10^{-27}$ cm$^3$s$^{-1}$ for an Einasto profile. While the wino and Higgsino line signals are at the right order of magnitude to produce this signal, the bino line signals fall several orders of magnitude short of this requirement. The required cross section can be lowered by allowing for a steeper dark matter profile at the Galactic Center -- for instance, \cite{130weniger} finds that a cross section of $\sim2\times 10^{-28}$ cm$^3$s$^{-1}$ can explain the signal close to the Galactic Center (Reg4 and Reg5 in \cite{130weniger}), where the signal is the most significant, for a contracted Navarro, Frenk, and White (NFW) dark matter profile with slope $\alpha=1.3$. This, however, is still insufficient to bring line signals from the bino into contention.   

The wino and Higgsino, meanwhile, annihilate dominantly into gauge bosons, and their subsequent decays produce a significant continuum of photons at lower energies. The size of this continuum is represented by  the ratio $R^{th}$, listed in the final column of Table \ref{tab:limits}, defined in \cite{continuum} as
\begin{equation}
R^{th}\equiv\frac{\sigma_{ann}}{2\sigma_{\gamma\gamma}+\sigma_{\gamma Z}},
\end{equation}
which is to be constrained through comparison with the analogous ratio from observation \cite{continuum}
\begin{equation}
R^{ob}\equiv\frac{1}{n^\gamma_{ann}}\frac{N_{ann}}{N_{\gamma\gamma}+N_{\gamma Z}},
\end{equation}
where N refers to the number of photons from the relevant process, and $n^\gamma_{ann}$ is the total number of photons per annihilation in the considered energy range. In \cite{continuum}, simply requiring that the continuum contribution not supersaturate the data was found to constrain $R^{ob}$ to be below 75 to 120\footnote{Allowing for a power law background makes this constraint even stronger \cite{continuum}.} for dark matter in the mass range 125 to 150 GeV annihilating primarily into W or Z bosons. Line contributions from wino or Higgsino dark matter are in clear tension with this bound, ruling them out as an explanation of the feature observed by Fermi. It should be noted that a 130\,GeV dark matter particle annihilating into Ws is also in tension with PAMELA antiproton constraints (see e.g. \cite{antiprotons},\cite{antiprotons2}) and observation of dwarf galaxies (\cite{fermidwarfs}; also see \cite{dwarf} for related uncertainties with dwarf galaxies).

Despite the incompatibility of a line signal, internal bremsstrahlung (IB) might, as motivated in the previous section, still salvage supersymmetry as an explanation of the 130 GeV signal observed by Fermi. IB refers to radiation of a photon from either the charged Standard Model (SM) final states that dark matter annihilates into or the charged mediator in the t- or u-channel. One can nominally distinguish between the two as final state radiation (FSR) and virtual internal bremsstrahlung (VIB) respectively, but they cannot be treated separately in a gauge-invariant manner and must always be considered and calculated together.  




The IB component from neutralino annihilation is known to be the most prominent when annihilation is into particles that are effectively massless relative to the neutralino, and the virtual particle that mediates the process is close in mass to the neutralino \cite{ibdiscussion}\cite{ibheavyneutralino}. Since the W and Z gauge bosons are massive final states for a neutralino of mass around 130-150\,GeV, IB from wino or Higgsino dark matter does not produce a feature sharp enough to explain the Fermi observation, despite the presence of a degenerate chargino to mediate its annihilation; this has been verified explicitly.  

IB from bino dark matter, on the other hand, is more promising. The main annihilation channels for a bino are to fermion pairs, mediated by the corresponding sfermions. For nonrelativistic annihilation in the halo, the cross section for this process is helicity suppressed by a factor of $(m_f/m_{\chi})^2$; since the top quark is heavier than the dark matter mass of interest here and all other SM fermions are $\mathcal{O}$(GeV) or lighter, this suppression is of several orders of magnitude, and acts as an efficient mechanism to suppress the continuum photon production. The addition of a photon in the final state, on the other hand, lifts this helicity suppression, and $\sigma v(\chi\chi\rightarrow\bar{f}f\gamma)$ can be comparable to $\sigma v(\chi\chi\rightarrow\bar{f}f)$. Since fermions and sfermions couple to the bino via hypercharge and leptons have larger hypercharge than quarks (also, sleptons are generally lighter than squarks),  IB primarily involves leptonic channels. 

For an almost pure bino, the IB cross section is fairly robust; in the limit of massless fermions, it is approximately given by \cite{ibdiscussion}
\begin{align}
\label{eq:ibspectrum}
\frac{d\sigma_{\chi\chi\rightarrow\bar{f}f\gamma}}{dx}&=\,\alpha_{EM}Q_f^2\frac{|\tilde{g}_L|^4+|\tilde{g}_R|^4}{64\pi^2m_\chi^2} (1-x)\nonumber\\
&\times\left(\frac{4x}{(1+\mu)(1+\mu-2x)}-\frac{2x}{(1+\mu-2x)}-\frac{(1+\mu)(1+\mu-2x)}{(1+\mu-x)^2}\text{log}\frac{1+\mu}{1+\mu-2x}\right)
\end{align}
where $\mu=(m_{\tilde{f_L}}/m_\chi)^2=(m_{\tilde{f_R}}/m_\chi)^2$ (assuming the same mass for both sfermions), $g_{L(R)}$ corresponds to the coupling of the left(right) handed sfermion to the bino, and $x=E_\gamma/m_\chi$. Note the prominent log enhancement close to the kinematic edge ($x\sim1$) when sfermions are approximately degenerate with the bino ($\mu\sim 1$); light sleptons are therefore a crucial element of strong IB features.   

This setup offers a clear strategy towards an attempt at a supersymmetric explanation of the Fermi signal. The primary contribution must come from IB from mostly bino dark matter in the 130 to 150 GeV range, with sleptons not too far above in mass to produce a sufficiently large and sharp feature. A small wino and/or higgsino component can augment this signal via a $\gamma\gamma$ or $\gamma Z$ peak as long as the production of continuum photons is sufficiently suppressed to evade the observational bounds. 


\section{Approach, Results, and Discussion}

The following analysis is based on Fermi LAT data from the inner 3$^\circ$ radius region around the Galactic Center, where the 130\,GeV signal was found to be the most significant \cite{130finkbeiner}\cite{130tempel}. For this purpose, event counts as listed in Appendix A of \cite{continuum} (unmasked region), corresponding to ULTRACLEAN events in the Pass 7\_Version 6 release\footnote{http://heasarc.gsfc.nasa.gov/FTP/fermi/data/lat/weekly/p7v6},  for 128 energy bins from 5.1\,GeV to 198\,GeV are used. The interested reader is referred to that paper for details of energy binning and individual photons counts. 

The gamma ray spectrum from neutralino annihilation is generated using DarkSUSY version 5.0.5\footnote{ A subtlety regarding the IB component from DarkSUSY is worth mentioning here. DarkSUSY obtains the spectrum from Pythia, which simulates the annihilation process as a decay of a hypothetical particle of mass $2m_\chi$, hence missing contributions corresponding to virtual IB. DarkSUSY corrects for this by subtracting the FSR spectrum of the hypothetical decay from the Pythia result and adding the full IB contribution \cite{ibdiscussion}. Spectra obtained from Pythia and DarkSUSY might show differences due to this correction. The author thanks Joakim Edsj\"{o} for clarification on this issue.} \cite{darksusy}. The spectrum is normalized such that $\sigma_{\gamma\gamma}v=2\times10^{-28}$cm$^3$s$^{-1}$ corresponds to 30.3 photons in the $\gamma\gamma$ peak. This choice is made for the following reason. The authors of \cite{continuum} find that the data set in question is best fit by a $\gamma\gamma$ peak with 30.3 photons. The lowest cross section into two photons that is consistent with the signal in this region, meanwhile, is roughly $\sigma_{\gamma\gamma}v=2\times10^{-28}$cm$^3$s$^{-1}$ (\cite{130weniger}, for Reg4 and Reg5 therein), for a contracted NFW dark matter profile with a modified slope $\alpha=1.3$. While an Einasto profile with $\sigma_{\gamma\gamma}v=1.27\times10^{-27}$cm$^3$s$^{-1}$ is more consistent with data from extended regions beyond the 3$^\circ$ region considered here \cite{130weniger} \footnote{See \cite{tcreview} for a detailed discussion of the compatibility of various profiles to the signal from extended regions. In particular, the NFW profile used here is consistent with the signal in the 3$^\circ$ region considered here but needs to be modified at larger angles to maintain consistency with data from extended regions. The author thanks Torsten Bringmann and Christoph Weniger for pointing this out.}, the number chosen above is a better reflection of what the cross section at least needs to be in order to be considered a viable option.

Corrections for the Instrument Response Function of the Fermi LAT, which describes the energy dispersion of incident photons, are crucial for line signals and the sharp IB feature at the kinematic edge, and less important for the continuum at lower energies. This is taken into account by fitting the energy dispersion at 130\,GeV, plotted in Appendix B of \cite{continuum}, with a Gaussian, and applying this dispersion correction to the dark matter spectrum above 100\,GeV. Correction for the change in effective area of the instrument at different energies is also approximately incorporated using the information provided in \cite{fermi_irf}. A more careful treatment of these factors, while possible, is unnecessary for the major objectives of this paper.  

Finally, in addition to the continuum from the decay of annihilation products, other potentially important dark matter contributions to the gamma ray spectrum also need to be considered. Of these, Inverse Compton Scattering (ICS) of photons in the interstellar radiation field off charged products from dark matter annihilation is the most important.  The ICS contribution is estimated using a semi-analytic formalism described in \cite{positrons},\cite{patrick}, with the simplification of ignoring spatial diffusion (equivalent to setting the halo function to unity),  and found to be negligible; this is understandable, since the production of fermions is strongly helicity suppressed.

\subsection{Scan Results}

A scan over the MSSM parameter space, optimized for bino dark matter and the Fermi signal, was performed with DarkSUSY. The lightest neutralino was required to have a mass between 120\,GeV and 160\,GeV, and be mostly bino, with the wino/Higgsino components required to be sufficiently suppressed to avoid overproduction of the continuum. A single mass value was chosen for all sleptons, and this value was constrained to be within 20\,GeV of the lightest neutralino mass. Squark masses were fixed at 1\,TeV. The bino mass $M_1$ was varied between 120 and 160\,GeV, and the wino mass $M_2$ and the higgsino mass parameter $\mu$ were varied between 120\,GeV and 2\,TeV; all three were allowed to take negative values. 


In the Fermi data set used in this analysis, the excess appears at energies between 121.62\,GeV and 136.40\,GeV,  where a total of 24 photons are observed (Appendix A in \cite{continuum}). The first issue of concern is whether the signal from neutralino annihilation is strong enough to explain this excess. A falling power law background, obtained from a fit to the spectrum at lower energies, contributes 6 or 7 photons in this energy range. A good fit should be possible with $\sim$9 or more photons from neutralino annihilation in addition to this background contribution, as the observed count of 24 is then 2$\sigma$ or less away (assuming $\sigma\sim\sqrt{N}$). 
For comparison, a $\gamma\gamma$ peak with 30.3 photons, the best monochromatic fit to data \cite{continuum}, contributes $\sim$19 photons in this energy range after energy dispersion. 

\begin{figure}[t]
\centering
\includegraphics[width=3.2in,height=2.2in]{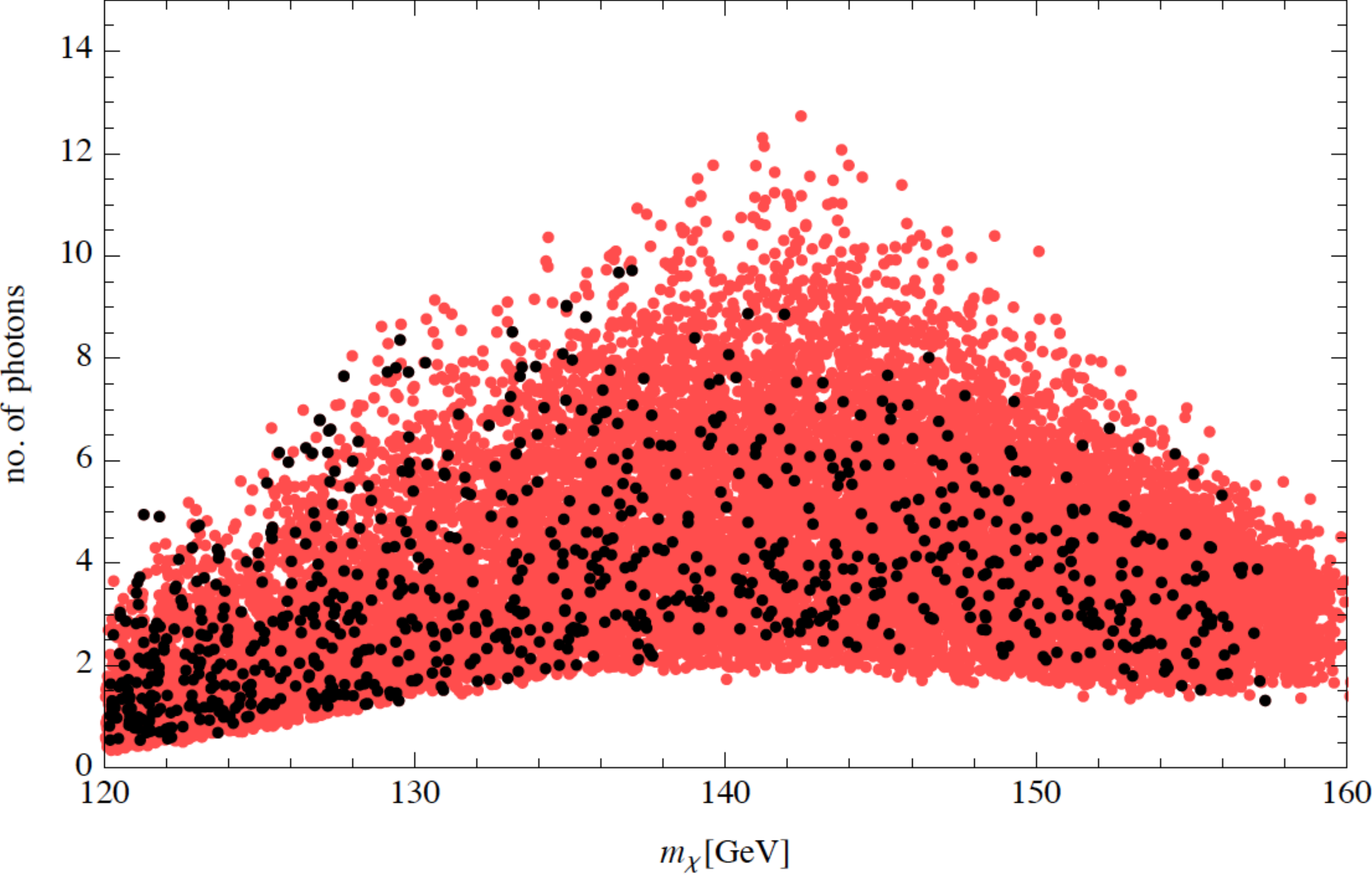}
\includegraphics[width=3.2in,height=2.2in]{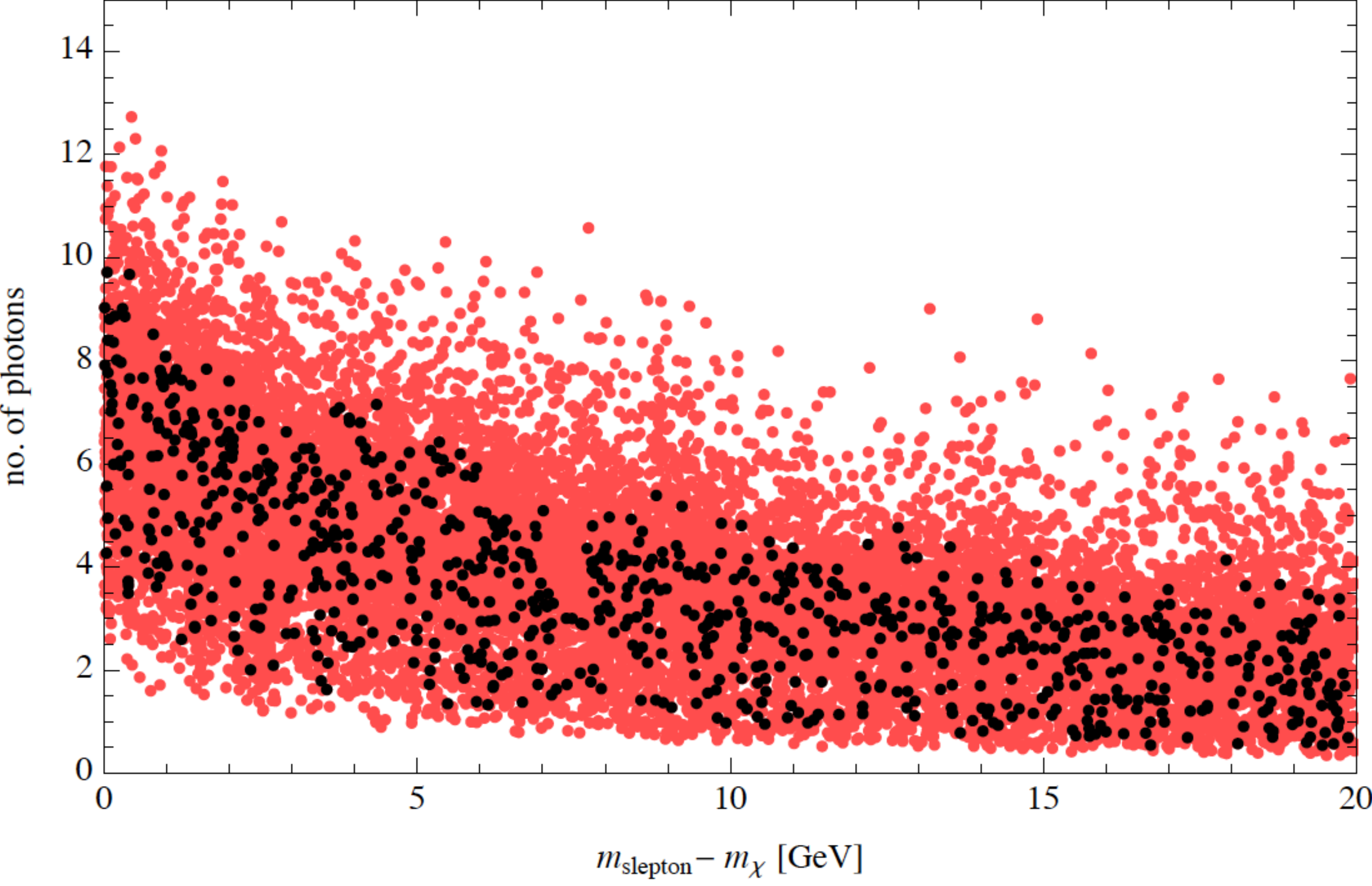}
\caption{The total number of photons in the 121.62\,GeV to 136.40\,GeV energy range from dark matter annihilation, as a function of dark matter mass (left) and slepton-neutralino mass difference (right), for a scan over primarily bino candidates. Points with thermal relic density calculated to be in the range $0.1\leq \Omega h^2\leq 0.124$, making them consistent with relic density constraints, are shown in black. }
\label{fig:photonnumber}
\end{figure}

The first plot in Figure \ref{fig:photonnumber} shows the number of photons in this energy range between 121.62\,GeV and 136.40\,GeV from dark matter annihilation as a function of dark matter mass.  The number is fairly robust in the range of dark matter masses that can explain the Fermi signal: dark matter contributes $\mathcal{O}$(few) photons, peaking at $\sim$ 13 photons around $m_\chi\approx 145\,$GeV. The occurrence of the peak at this energy is understandable: for $m_\chi\approx145$\,GeV, the $\gamma Z$ channel gives monoenergetic photons at $E_\gamma=m_\chi (1-m_Z^2/4m_\chi^2)\approx130$\,GeV, which can be sizable even with a tiny wino/Higgsino component. Meanwhile, the peak of the IB feature still falls mostly into the 121.62\,GeV to 136.40\,GeV range (recall that the best fit for a purely IB signal occurs for $m_\chi\sim150$\,GeV); hence both IB and the $\gamma Z$ line are at ideal energies to contribute towards the signal. As the mass changes away from this ideal value, either the IB peak or the line contribution is lost, and the signal dies away, as is seen in the plot. 

The second plot in Figure \ref{fig:photonnumber} shows the dependence on the slepton-neutralino mass difference, which follows what is expected from Equation \ref{eq:ibspectrum}; the photon count gradually rises as the sleptons become more degenerate in mass with the neutralino. The scan is terminated at 20\,GeV, as the downward trend continues and a sufficient number of photons cannot be obtained.   

The results of this scan show that, with a combination of sleptons approximately degenerate with the dark matter particle and contributions from both IB and line components, a dark matter signal large enough to explain the Fermi feature can be possible without overproduction of the continuum. Attention must now switch to whether the contribution is of the right shape to explain the observed signal. 

\subsection{Benchmark Points and Fit to Data}

This section discusses four benchmark points (labelled BM1, BM2, BM3, and BM4) that are representative of the scanned sample, and their fits to the Fermi signal. These points are listed in Table \ref{tab:bm}, with other relevant information. It should be stressed that these were chosen to highlight distinct features of signals that are possible with internal bremsstrahlung, and are not the points that best fit the data. 

\begin{table}[h]
\begin{center}
\begin{tabular}{|c|c|c|c|c|} \hline
~ & ~~~$BM1$~~~&~~~$BM2$~~~ & ~~~$BM3$~~~&~~~$BM4$~~~\\
\hline
$M_1$ & 135.2 & 144.7  &145.6 & 138.2\\
$M_2$& 235.5 & 152.8 &150.4 & 161.2\\
$\mu$ & -489.9 & 838.4 & 783.0 & 512.9\\
tan$\beta$& 18.5 & 6.6 & 33.2 & 20.5\\
$m_{\tilde{l}}$& 136.7 & 156.6 & 146.7 & 138.5\\
\hline
$m_\chi$ & 134.4 & 143.0 & 144.7 & 136.4\\
bino fraction & 0.99 & .90 & 0.91 & 0.97\\
$\Omega h^2$ & 0.19 & .0058 & 0.0033 & 0.11\\
$n_\gamma$ from IB & 4.8 & 1.8 & 4.5 & 14.7\\
$n_\gamma$ ($\gamma\gamma +\gamma Z$) & 2.0 & 5.1 & 5.2 & 4.4\\
TS & 15.8 & -- & -- & 17.8\\
significance & 4.0 & -- & -- & 4.2\\
\hline
\end{tabular}
\caption{Four benchmark points chosen for detailed study and fit to Fermi data, and fit results. All masses are in GeV. $n_\gamma$ refers to the number of photons contributed to the 121.62\,GeV to 136.40\,GeV energy bin.}
\label{tab:bm}
\end{center}
\end{table}

To perform the fit to Fermi data, the spectrum generated with DarkSUSY for each benchmark point is added to a single falling power law background, with the normalization and the power law index allowed to vary to give the best fit. Such fits are not possible for points for which the low energy continuum is close to saturating the Fermi data; however, such points are still useful to illustrate various characteristics of IB signals. This is true of benchmark points BM2 and BM3. Therefore, in this section, fits with background are only performed where feasible (BM1 and BM4). It should be noted, however, that the assumption of a single power law background across the entire spectrum is a rather strong one, and signals not allowed by such a background might be consistent with other forms of background.  


Following \cite{brem},\cite{continuum}, the significance of the fit is estimated by maximizing the likelihood function 
\begin{equation}
\ln{\mathcal{L}}=\sum_{k=1}^{N_{bins}} n_k\cdot \ln\phi_k-\phi_k-\ln(n_k!)
\end{equation}
where $n_k$ ($\phi_k$) represents the number of photons observed (expected) in the $k^{th}$ energy bin. This is then used to calculate the test statistics (TS)
\begin{equation}
TS=-2\ln\frac{\mathcal{L}_{\text{null}}}{\mathcal{L}_{\text{benchmark}}}
\end{equation}
where $\mathcal{L}_{\text{benchmark}}$ is the likelihood of the benchmark model, and $\mathcal{L}_{\text{null}}$ corresponds to the null hypothesis, i.e. fit with a power law background only. The nominal significance of the fit is taken to be $\sqrt{TS}$. 

\begin{figure}[t]
\centering
\includegraphics[width=3.2in,height=2in]{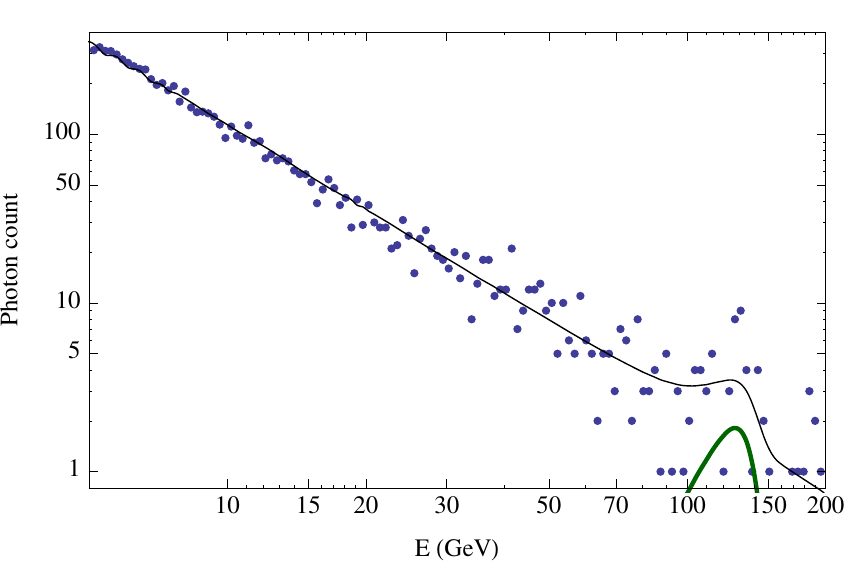}
\includegraphics[width=3.2in,height=2in]{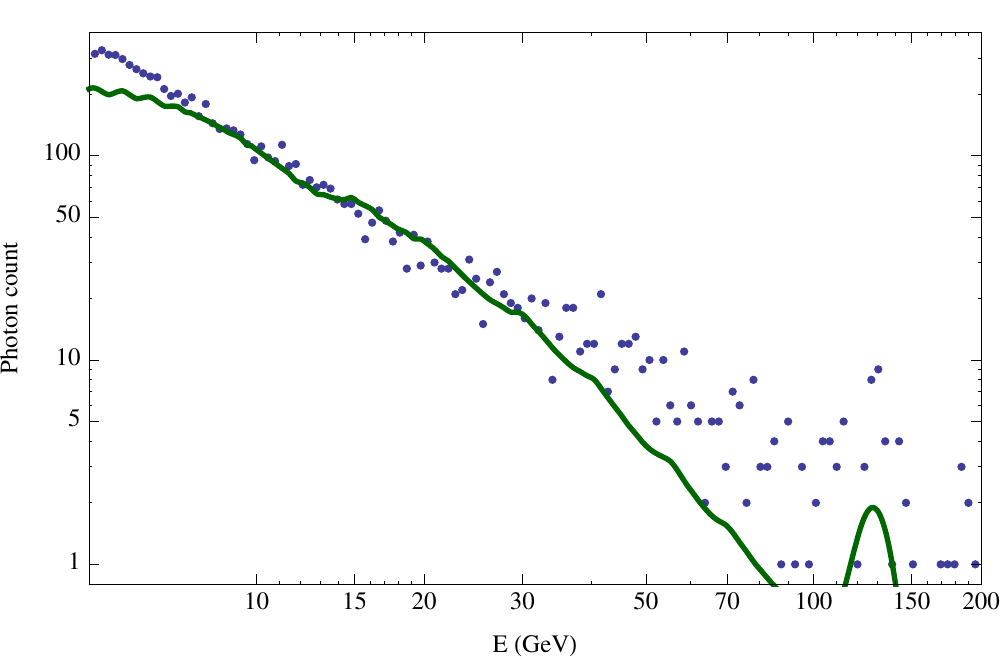}\\
~\\
\includegraphics[width=3.2in,height=2in]{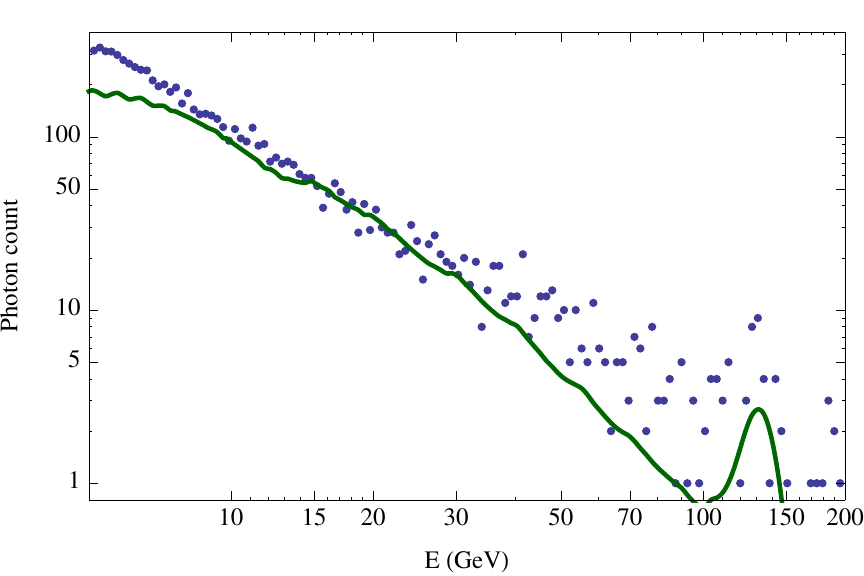}
\includegraphics[width=3.2in,height=2in]{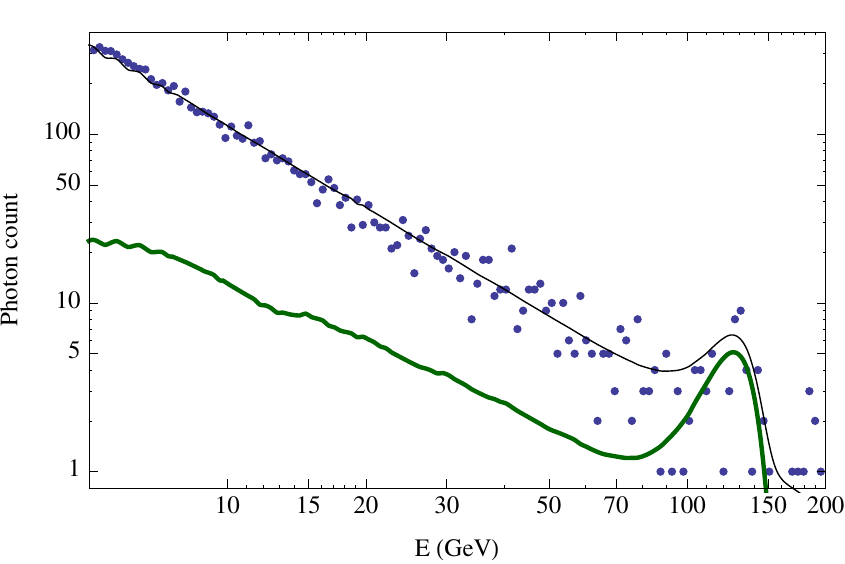}\\
\caption{Fermi data from the inner 3$^{\circ}$ of the Galactic Center  for all 128 energy bins (blue dots, as listed in Appendix A of \cite{continuum}) and the gamma ray spectra from dark matter (green) for the four benchmark points BM1, BM2 (top row), BM3, and BM4 (bottom row). The black curves for BM1 and BM4 represent the overall fit to the signal, consisting of a single power law background in addition to the dark matter signal; BM2 and BM3 supersaturate the continuum at lower energies and do not allow such fits.
}
\label{fig:fits1}
\end{figure}

The gamma ray spectra from each of the benchmark points, superimposed on the Fermi data over the entire energy range between 5 and 198 GeV, are plotted in Figure \ref{fig:fits1}. Information regarding the significance of the fit is listed in Table \ref{tab:bm}. As mentioned earlier, the fit is only performed for two of the benchmark points, BM1 and BM4; the other two benchmark points contain large continuum signals at low energies that saturate the Fermi data and are therefore incompatible with a single power law background, making such fits impossible. The reader is advised to use caution in interpreting these fit results, since these are not produced from an extensive scan and detailed fitting procedure, which is not the main purpose of this paper, and are merely meant to be a rough indication of  the compatibility between prediction and signal. 



Next, each benchmark point and its fit to data is discussed in turn. 

BM1 is an almost pure bino that contributes dominantly via IB. The mass degeneracy between the neutralino and the sleptons --  the difference is only 2\,GeV -- makes the IB feature very sharp and prominent above the rest of the spectrum. This very mass degeneracy also facilitates coannihlations in the early universe, leading to a relic density very close to the observed value. It can be seen that the pure bino comes with an extremely efficient suppression of continuum photons -- the continuum is so small that it does not even appear on Figure \ref{fig:fits1} -- due to helicity suppression of annihilation into fermions. However, it also comes with the disadvantage that the desired number of photons is difficult to obtain via a purely IB contribution, even with extremely degenerate sleptons.

BM2 improves on BM1 by introducing a 10\% wino admixture to the neutralino, opening up significant contributions from the $\gamma Z$ line: in this case, the line contributes 5 photons to the signal bin. On the other hand, this wino contribution also floods the spectrum with continuum photons at lower energies that, for this particular benchmark point, saturate the Fermi signal in the region around $15-20$\,GeV (Figure \ref{fig:fits1}), making the signal incompatible with a single power law background. The thermal relic density also plummets due to extremely efficient annihilation through the wino component. The sleptons are about 15\,GeV heavier than the neutralino, and the lifting of this mass degeneracy suppresses the IB contribution relative to BM1, so that the line signal is the primary component contributing to the 130\,GeV signal. Although BM1 and BM2 contain essentially the same number of photons in the signal bin, the line provides a narrower feature that better fits the shape of the Fermi signal around 130\,GeV.

BM3 combines the virtues of both BM1 and BM2: sleptons degenerate with the neutralino lead to a sharp IB feature, while a small wino admixture contributes a prominent $\gamma Z$ line signal, resulting in a relative abundance of photons in the right energy bin. Such points, characterized by sleptons approximately degenerate with the neutralino as well as line contribution from a small wino component, should be considered the most promising avenues for producing the observed signal. As with BM2, dark matter contribution saturates the Fermi signal in the $15-20$\,GeV window, making the signal incompatible with a single power law background, and the thermal relic density is smaller than desired due to the wino component. 

BM4 analyses a fit with about 19 photons -- the same number as from a best fit monochromatic line -- in the 121.62\,GeV to 136.40\,GeV energy bin. This is achieved by enhancing the dark matter signal by an additional factor of 3, and could correspond to some astrophysical enhancement such as a steeper dark matter profile at the Galactic Center or substructure along the line of sight towards this region. A significantly better fit is obtained in this case, and the enhancement allows the signal to be composed mostly of IB, enabling further suppression of the wino component relative to BM2 or BM3. This suppression has two advantages: one, the continuum is suppressed relative to BM2 and BM3, allowing for a good fit with a single power law background; two, the relic density is raised to a value that matches observation.  

That the subdominant admixtures in all of BM2, BM3, and BM4 are winos is not a coincidence. There are several reasons for the wino being preferred to Higgsinos. Because Higgsinos fare far worse with continuum photons than winos (see Table \ref{tab:limits}), they can contribute no more than 1 or 2 photons to the 130\,GeV signal before saturating the spectrum at lower energies. In addition, since the Higgsino annihilation cross sections -- both total and into lines -- are smaller than the corresponding ones for a wino by an order of magnitude, producing the same number of line photons as the wino requires a larger Higgsino fraction, suppressing the bino fraction and consequently the IB contribution.  Finally, the $\gamma Z$ line is more prominent than $\gamma\gamma$ for a wino, which is desirable.

The four benchmark points discussed in this subsection each illustrate an important advantage --  suppression of continuum photons, elimination of the need for sleptons within a few GeV of the lightest neutralino, maximization of photon count in the bins where Fermi observes a signal, and significant improvements from additional $\mathcal{O}(1)$ boosts, respectively -- but demand caution in other aspects. The above analysis serves to highlight the interplay of the various factors that are relevant for a supersymmetric explanation of the Fermi signal, but confirms nevertheless that such an explanation is indeed possible with IB.

\section{Other Considerations}

This section is devoted to brief discussions of various related aspects that deserve attention.

\subsection{ Relic Density}

The bino, being a gauge singlet, is generally inert, leading to a thermal relic density that, for a bino of mass around 130\,GeV, is far in excess of the observed value. However, there exist well-understood ways to resolve this discrepancy. One is to have an almost degenerate slepton (usually stau), within $\sim 5$\% of the dark matter mass, to enable efficient coannihilation \cite{staucoannihilation}; another is to make the neutralino well-tempered, i.e., introduce a small admixture of wino or Higgsino \cite{welltempered}. Both of these features are prominent in the class of models discussed in this paper (recall that the former led to a reasonably good relic density for BM1, while both factors were at play in obtaining the correct relic density for BM4), and obtaining a thermal relic density in agreement with observations appears to be a tractable task. For the set of points scanned over with DarkSUSY, the computed relic densities ranged from $0.001\leq \Omega h^2\leq 0.3$, with a significant fraction within $2\sigma$ of the current best fit value; recall that these points that are in agreement with observation were plotted in black in Figure \ref{fig:photonnumber}.   

\subsection{Collider and Direct Detection Constraints}

Collider and direct detection searches generally place stringent constraints on dark matter and supersymmetry, and must be considered. For mostly bino dark matter and squarks at the TeV scale or heavier, Tevatron and LHC constraints are easily avoided. Despite the candidate considered here having appreciable couplings to leptons, facilitated by the presence of light sleptons, LEP constraints on dark matter\cite{lep} only apply to dark matter masses below its threshold of 100\,GeV, and are irrelevant to the mass range of interest in this paper. Likewise, when squarks are heavy and the lightest neutralino is a sufficiently pure gaugino, tree level spin-independent direct detection interactions with nuclei are suppressed, and the candidate is safe from the direct detection bounds placed by XENON100 \cite{xenon100} (see \cite{ftsusy},\cite{ftsusy2} for elaboration and a detailed study of this point). The dark matter candidates studied in this paper are therefore safe from both collider and direct detection constraints at present.

\subsection{Astrophysical Uncertainties}

While gamma rays represent the cleanest indirect detection channels for dark matter, a dark matter interpretation of the Fermi 130\,GeV signal is still plagued with astrophysical uncertainties. Of these, a precise knowledge of the astrophysical background --  even at the level of whether it follows a single power law across the entire energy range of interest -- and reliable knowledge of the dark matter profile at the Galactic Center, or the existence of substructures in the direction of the signal, introduce huge uncertainties in translating observations to implications for possible underlying particle physics models. For instance, if dark matter at the Galactic Center follows an isothermal profile, the combination of IB and line signals presented here are no longer plausible as an explanation of the Fermi signal. On the other hand, even a modest presence of dark matter substructure in the direction of the inner $3^\circ$ of the Galactic Center can greatly enhance a dark matter signal, allowing a larger portion of supersymmetric parameter space --  such as purely IB contributions, or IB with heavier sleptons --  into consideration, or enables better fits\footnote{BM4 serves as an illustration of this point.}, even with less peaked dark matter profiles such as the more favored Einasto profile. It should therefore always be kept in mind that the uncertainties introduced by astrophysical factors are large and can have significant implications. 

\subsection{Naturalness}

With a relatively heavy Higgs discovered and the LHC failing to find any light superpartners, the naturalness of supersymmetry as a resolution of the hierarchy problem has become an important issue. A commonly employed measure of fine-tuning in terms of the tree level Z boson mass requires $\mu\sim m_Z$, with the amount of fine-tuning scaling as $\sim (\mu/m_Z)^2$ (see e.g.\cite{golden},\cite{ftsusy} for more detailed discussions). The scenario presented in this paper, where the neutralino is almost entirely gaugino, and the $\mu$ parameter is required to be extremely large in order to suppress continuum contributions from the Higgsino component, is significantly fine-tuned in this regard. This, however, is strictly true only in the MSSM, where the fine-tuning problem is already known to be serious. In contrast, in nonminimal versions of supersymmetry favored by a 125\,GeV Higgs\cite{HRP}, such as the Next-to-Minimal Supersymmetric Standard Model (NMSSM) or $\lambda$-SUSY, a parametric suppression of fine-tuning can occur, and larger values of $\mu$ can be perfectly natural (see \cite{ftsusy2} for a detailed discussion). Although the discussion presented here was confined to the MSSM, a neutralino that is mostly bino with a small wino component can be easily realized in such nonminimal extensions, and would be consistent with naturalness.

\section{Conclusions}

There is now clear evidence of an unexplained feature at 130\,GeV towards the Galactic Center in the Fermi LAT data. A dark matter interpretation is very tempting, and the feature fits extremely well to dark matter annihilating into monochromatic photons, the long anticipated ``smoking gun" signature of dark matter. The most studied dark matter candidate, the lightest neutralino in supersymmetry, is incompatible with this line interpretation of the signal, constrained by the absence of a continuum at low energies in the observed data, or small cross sections into line photons. 

The purpose of this paper was to examine whether these constraints can be circumvented in supersymmetric scenarios where internal bremsstrahlung plays a prominent role. Satisfactory scenarios were indeed found. A few benchmark models illustrating the major possibilities were presented, and their agreement with Fermi data explored.  

The scenario most consistent with the 130\,GeV signal corresponds to sharp internal bremsstrahlung from a $\sim 145$\,GeV mostly bino dark matter particle in conjunction with a $\gamma Z$ line from a subdominant wino component; with this choice of mass, the peaks of both the IB and $\gamma Z$ signals fall in the 130\,GeV region, producing a strong signal. Light sleptons approximately degenerate with the neutralino are required to make the IB feature prominent. This combination of bino dominance, approximate mass degeneracy of the sleptons and the neutralino, and a possible line contribution from a subdominant wino component is a generic feature of the class of candidates studied in this paper. The presence of light sleptons also facilitates coannihilations, providing thermal relic densities roughly in agreement with observation. 

A contracted NFW profile was chosen over the more favored Einasto profile to allow for a more generous -- therefore broader -- treatment of the parameter space; the photon flux from choosing the latter profile is only an $\mathcal{O}$(1) factor smaller, and a modest contribution from, for instance, substructure along the line of sight to the center of the galaxy can easily overcome this difference. Given the large uncertainties in these astrophysical factors, possibilities of such corrections should not be ignored.

While instrumental or nonconventional astrophysical effects might yet explain this 130\,GeV anomaly, the possibility that this might be the first signature of dark matter -- of a particle beyond the Standard Model -- is one with tremendous implications, and one worth pursuing even in the midst of uncertainty. More data, from the Galactic Center and elsewhere, and with Fermi as well as with other instruments, will gradually improve the details of the signal, leading to a clearer picture. For the moment, the possibility that the signal has its origins in dark matter annihilation remains alive; this paper has presented a case that so too does the possibility that that origin is supersymmetric.

\vskip0.8cm
\noindent{\large \bf Note Added} 
\vskip0.3cm
\noindent During the completion of this project, a review paper \cite{tcreview} appeared, where an MSSM scan exploring the relative sizes of the IB, line signals, and secondary photons in the context of the Fermi 130\,GeV signal is presented and discussed. The results presented in this paper are in agreement with the results therein.

\vskip0.8cm
\noindent{\large \bf Acknowledgements} 
\vskip0.3cm

\noindent The author sincerely thanks Mariangela Lisanti, Maxim Perelstein, and Tracy Slatyer for insightful discussions and constructive comments on the manuscript. The author also acknowledges helpful discussions with Torsten Bringmann, Rouven Essig, Patrick Meade, and Christoph Weniger. This research is supported by the U.S. National Science Foundation through grant PHY-0757868.



\end{document}